\documentclass[11pt,superscriptaddress,aps,prd,preprint,showpacs,nofootinbib]{revtex4}
\usepackage[dvips]{graphicx}
\usepackage{amsfonts}
\usepackage{slashed}

\newcommand{\be}{\begin{equation}}
\newcommand{\en}{\end{equation}}
\newcommand{\ba}{\begin{eqnarray}}
\newcommand{\ea}{\end{eqnarray}}
\newcommand{\bea}{\begin{eqnarray}}
\newcommand{\eea}{\end{eqnarray}}

\begin{document}

\title{Causal G\"{o}del-type metrics in non-local gravity theories}

\author{J. R. Nascimento}
\email[]{jroberto@fisica.ufpb.br}
\affiliation{Departamento de F\'{\i}sica, Universidade Federal da 
Para\'{\i}ba,\\
 Caixa Postal 5008, 58051-970, Jo\~ao Pessoa, Para\'{\i}ba, Brazil}

\author{A. Yu. Petrov}
\email[]{petrov@fisica.ufpb.br}
\affiliation{Departamento de F\'{\i}sica, Universidade Federal da Para\'{\i}ba,\\
 Caixa Postal 5008, 58051-970, Jo\~ao Pessoa, Para\'{\i}ba, Brazil}

\author{P. J. Porf\'{\i}rio}
\email[]{pporfirio@fisica.ufpb.br}
\affiliation{Departamento de F\'{\i}sica, Universidade Federal da 
Para\'{\i}ba,\\
 Caixa Postal 5008, 58051-970, Jo\~ao Pessoa, Para\'{\i}ba, Brazil}


\begin{abstract}
It is well known that non-local theories of gravity have been a flourish arena of studies for many reasons, for instance, the UV incompleteness of General Relativity (GR). In this paper we check the consistency of ST-homogeneous G\"{o}del-type metrics within the non-local gravity framework. The non-local models considered here are ghost-free but not necessarily renormalizable since we focus on the classical solutions of the field equations. Furthermore, the non-locality is displayed in the action through transcendental entire functions of the d'Alembert operator $\Box$ that are mathematically represented by a power series of the $\Box$ operator. We find two exact solutions for the field equations correspondent to the degenerate ($\omega=0$) and hyperbolic ($m^{2}=4\omega^2$) classes of ST-homogeneous G\"{o}del-type metrics. 

\end{abstract}


\maketitle

\section{Introduction}
\label{sec:intro}

Famous observational results of recent years, such as discovery of accelerated expansion of the Universe \cite{Riess}, observation of gravitational waves \cite{LIGO} and of black holes \cite{Akiyama},  increased strongly the interest to theoretical studies of gravity aimed to solve two main problems -- first, explanation of cosmic acceleration, and second, development of a consistent perturbative description for gravity. While actually there are various models explaining the late-time accelerated expansion of the Universe, based on extension of gravity sector or adding extra fields (for a general review on modified gravity see \cite{mybook}), the problem of quantum description of gravity is much more complicated -- the usual Einstein gravity is known to be non-renormalizable, higher-order curvature counterterms must be included in the action in order to improve the UV perturbative behavior of the theory. In the meantime, it implies in arising of ghosts, that is, negative-norm states. One of the most promising ways to solve the problem of ghosts is based on the development of a nonlocal generalization for gravity which could guarantee renormalizability or even UV finiteness for a corresponding theory. Originally, the non-locality has been proposed  in various contexts. For example, the first non-local models were studied in the \mbox{1950}'s \cite{Pais, Pauli:1953fsg, Chushiro}. On the other hand, many studies have been made within the phenomenological context in order to describe finite-size objects  at the quantum level, see f.e. \cite{Dirac:1962dmi, Efimov}, and further, the space-time nonlocality was treated as a very convenient solution of the problem of UV divergences which are ruled out due to form factors implying UV finiteness of many theories. Another motivation stems from the  string theory context where infinite-derivative operators naturally arise as a result of $\alpha^{\prime}$ (inverse of the string tension) corrections \cite{Kaku:1974zz,Green:1984fu,Witten:1985cc}. These aforementioned essential issues have naturally called attention of the mainstream and, as a result, many studies of various non-local extensions of gravity, both from classical and quantum viewpoints, have became increasingly popular in the literature. An interesting discussion of quantum aspects of nonlocal gravity is presented in \cite{Modesto0,Modesto05} (for a general review on nonlocal gravity, see also \cite{Modestorev}). At the same time, in this paper we intend to consider only classical issues.

 The paradigmatic example of the nonlocal gravity model is the theory proposed in \cite{Biswas1}. In this paper, the simplest nonlocal extension of gravity with the action $S=\frac{M^2_p}{2}\int d^4x\sqrt{-g}[R-\frac{1}{6}R(\frac{e^{-\Box/M^2}-1}{\Box})R]$ representing itself as a nonlocal generalization of known $R^2$-gravity has been introduced, and cosmological solutions in this theory were obtained. 

As it is known, one of the main tasks within classical studies of various modified gravity models is the testing of consistency of known solutions of GR in these theories. Within the nonlocal gravity, absolute majority of studies performed up to now only treated various aspects of cosmological solutions (among the most important papers devoted to this issue, we can indicate \cite{Odintsov2}, where cosmological solutions in a nonlocal model involving a function of $\Box^{-1}R$ were found, \cite{Biswas2}, where bouncing and inflating solutions were discussed, and \cite{Koshelev} where a very unusual nonlocal square-root gravity was proposed and proved to result in a hyper-exponential expansion). Some of the few papers dealing with other metrics within the nonlocal context are \cite{Modesto1,Kili} where also various solutions with a constant and vanishing scalar curvature, namely, the (anti)de Sitter, the simple 
G\"{o}del one, and $pp$-waves, were considered (some studies of perturbed Schwarzschild metric in a special form of a nonlocal gravity action are presented also in \cite{Mitra}). At the same time, it is clear that both other known metrics and more generalized gravity Lagrangians deserve detailed studies within the framework of nonlocal gravity.  So, in this paper we consider a more generic nonlocal gravity involving not only the scalar curvature but also contractions of Ricci and Weyl tensors, and a more broad class of metrics, that is, the G\"{o}del-type metrics characterized by two constant parameters $m$ and $\omega$, and, depending on these parameters, displaying either causal or non-causal behavior. 

As a result of the local homogeneity of the spacetime, all scalar invariants constructed on the basis of the Riemann tensor and its derivatives are constants. The manifolds satisfying this property are called Constant Scalar Invariants (CSI) spaces \cite{Coley:2005sq,Coley:2008th}. In particular, G\"{o}del-type metrics are  examples of CSI spaces, with their scalar invariants are entirely characterized by the metric parameters $m$ and $\omega$. One more again by virtue of the local homogeneity it is possible to pick a particular frame\footnote{In this work we will employ the Newman-Penrose (NP) formalism which consists of choosing a complex null frame \cite{penrose1984spinors,chandrasekhar1998mathematical}.} where field equations reduce to a set of algebraic ones. We will look for exact solutions and scrutinize the possibility of whether or not such solutions resolve the causality violation problem (an excellent review on the causality violation is presented in \cite{Luminet}).  

The structure of this paper looks like follows. In the section 2 we define the action of nonlocal gravity which will be studied in the paper. In the section 3 we review the main properties of the G\"{o}del-type metric. In the section 4 we discuss consistency of the G\"{o}del-type metric within our nonlocal gravity model. In the section 5 we present our conclusions. Finally, the Appendix is devoted to description of the boost-weight decomposition we use in the paper.

%
\section{Non-local gravity theory}
\label{secII}
%

We start this section introducing a generic non-local gravity theory proposed in \cite{Biswas:2013cha}. The action of the model reads
\begin{eqnarray}
\label{actmod}
\nonumber S&=&\frac{1}{2\kappa^2}\int d^{4}x\sqrt{-g}\, \left[R-2\Lambda_{\mbox{cc}}+\frac{1}{M^{2}_{N}}\left(R\mathcal{F}_{1}(\square)R+R_{\mu\nu}\mathcal{F}_{2}(\square)R^{\mu\nu}+C_{\mu\nu\alpha\beta\gamma}\mathcal{F}_{3}(\square)C^{\mu\nu\alpha\beta}\right)\right]\\
&+&S_{m}[g_{\mu\nu},\Psi],
\label{ac}
\end{eqnarray}   
where $\kappa^2$ is the gravitational constant, $\Lambda_{\mbox{cc}}$ is the cosmological constant and  the $M_{N}$ is a typical high energy scale where the non-locality effects must be important. On the other hand, the infrared (IR) regime is achieved by taking $M_{N}\rightarrow \infty$ where the theory recovers GR as it is naturally expected for any effective theory. We impose that the matter sources only couple to the metric in accordance with the equivalence principle which is encoded in the matter action $S_{m}[g_{\mu\nu},\Psi]$. Here, the $C_{\mu\nu\alpha\beta}$ is the Weyl tensor, $R_{\mu\nu}$ is the Ricci tensor and $R$ is the Ricci scalar. The functions $\mathcal{F}_{i's}(\square)$ are called form factor functions. It is worth stressing out that they must be transcendental (non-polynomial) entire functions of the d'Alembert operator $\Box$ in order to preclude the arising of non-physical degrees of freedom in the propagator \cite{Tomboulis:1997gg,Biswas:2005qr,Biswas:2011ar}. This by itself ensures the unitarity (absence of ghosts) of the theory, though further conditions must be taken into account to guarantee also the renormalizability of the theory \cite{Modesto1}. In this work we shall focus on the classical solutions thereby we are not concerned with the renormalizability conditions.

 From the mathematical point of view, an entire function is represented as a power series of its argument, in our specific case we can write the form factors as power series of the d'Alembert operator, i.e.,
\begin{equation}
\mathcal{F}_{i}(\square)=\sum_{n=0}^{\infty}f_{i_{n}}\square^{n}_{M_{N}},
\label{fr}
\end{equation}
where $\square_{M_{N}}\equiv\frac{\square}{M_{N}^{2}}$ is dimensionless and $f_{i_{n}}$ are the coefficients (dimensionless) of the power series in $\square_{M_{N}}$. Note that as it said before, the IR regime is achieved when the factor $\frac{\square}{M_{N}^{2}}$ is highly suppressed, i.e., $k\rightarrow 0$ (momentum space) and/or $M_{N}\rightarrow \infty$. On the other hand, the UV limit is hit when  $\mathcal{O}\left(\frac{\square}{M_{N}^{2}}\right)\propto 1$. 

Other choices for form factors have been worked out as well. For instance, in \cite{Biswas:2005qr,Biswas:2011ar} the form factors were picked to be
\begin{equation}
\mathcal{F}_{2}(\square)=-2\mathcal{F}_{1}(\square)=-\frac{1-e^{\square_{M_{N}}}}{\square_{M_{N}}}, \quad \mathcal{F}_{3}(\square)=0,
\end{equation}
of course leading to a full unitary (ghost-free) theory. Another possible choice of form factors leading to a ghost-free and also a renormalizable theory is
\begin{equation}
\mathcal{F}_{1}(\square)=-\frac{1}{3}\mathcal{F}_{3}(\square), \quad \mathcal{F}_{3}(\square)=\frac{e^{H(-\square_{M_{N}})}-1}{2\square_{M_{N}}},
\end{equation}
where the function $H(-\square_{M_{N}})=H(z)$ is defined in \cite{Tomboulis:1997gg} as:
\begin{equation}
e^{H(z)}=e^{\frac{a}{2}\left(\Gamma(0,p(z)^2)+\gamma_{E}+log(p(z)^2)\right)},
\end{equation}
where $\Gamma(0,p(z)^2)$ is the incomplete gamma function and $p(z)$ is a polynomial satisfying $p(0)=0$. Also, in \cite{Odintsov1} the non-local factors proportional to $\log\square$ were considered, being motivated by perturbative calculations.

\subsection{Field equations}

Varying the action (\ref{actmod}) with respect to the metric, one can obtain the following field equations \cite{Biswas:2013cha}:
\begin{eqnarray}
\nonumber E^{\alpha\beta}&=&G^{\alpha\beta}+\Lambda_{\mbox{cc}} g^{\alpha\beta}+P_{1}^{\alpha\beta}+P_{2}^{\alpha\beta}+P_{3}^{\alpha\beta}-2\Omega_{1}^{\alpha\beta}+g^{\alpha\beta}(g_{\mu\nu}\Omega^{\mu\nu}_{1}+\bar{\Omega}_{1})-2\Omega^{\alpha\beta}_{2}+\\
\nonumber&+&g^{\alpha\beta}(g_{\mu\nu}\Omega^{\mu\nu}_{2}+\bar{\Omega}_{2})-4\Delta^{\alpha\beta}_{2}-2\Omega^{\alpha\beta}_{3}+g^{\alpha\beta}(g_{\mu\nu}\Omega^{\mu\nu}_{3}+\bar{\Omega}_{3})-8\Delta^{\alpha\beta}_{3}=\\
&=&\kappa^{2}T^{\alpha\beta},
\label{ac1}
\end{eqnarray} 
where the tensorial quantities are defined as \cite{Biswas:2013cha}:
 \begin{eqnarray}
\nonumber P_{1}^{\alpha\beta}&=&\frac{1}{2M_{N}^{2}}\left[\left(4G^{\alpha\beta}+g^{\alpha\beta}R-4(\nabla^{\alpha}\nabla^{\beta}-g^{\alpha\beta}\square)\right)\mathcal{F}_{1}(\square)R\right];\\
\nonumber P_{2}^{\alpha\beta}&=&\frac{1}{2M_{N}^{2}}\bigg[4R^{\alpha}_{\,\nu}\mathcal{F}_{2}(\square)R^{\nu\beta}-g^{\alpha\beta}R^{\mu\nu}\mathcal{F}_{2}(\square)R_{\mu\nu}-4\nabla_{\nu}\nabla^{\beta}(\mathcal{F}_{2}(\square)R^{\nu\alpha})+2\square(\mathcal{F}_{2}(\square)R^{\alpha\beta})+\\
\nonumber&+&2g^{\alpha\beta}\nabla_{\mu}\nabla_{\nu}(\mathcal{F}_{2}(\square)R^{\mu\nu})\bigg];\\
\nonumber P_{3}^{\alpha\beta}&=&\frac{1}{2M_{N}^{2}}\bigg[-g^{\alpha\beta}C^{\mu\nu\sigma\gamma}\mathcal{F}_{3}(\square)C_{\mu\nu\sigma\gamma}+4C^{\alpha}_{\,\,\mu\nu\sigma}\mathcal{F}_{3}(\square)C^{\beta\mu\nu\sigma}-4(R_{\mu\nu}+2\nabla_{\mu}\nabla_{\nu})(\mathcal{F}_{3}(\square)C^{\beta\mu\nu\alpha})\bigg];\\
\nonumber\Omega^{\alpha\beta}_{1}&=&\frac{1}{2M_{N}^{2}}\sum_{n=1}^{\infty}f_{1_{n}}\sum_{l=0}^{n-1}\nabla^{\alpha}R^{(l)}\nabla^{\beta}R^{(n-l-1)}, \quad \bar{\Omega}_{1}=\frac{1}{2M_{N}^{2}}\sum_{n=1}^{\infty}f_{1_{n}}\sum_{l=0}^{n-1}R^{(l)}R^{(n-l)};\\
\nonumber\Omega^{\alpha\beta}_{2}&=&\frac{1}{2M_{N}^{2}}\sum_{n=1}^{\infty}f_{2_{n}}\sum_{l=0}^{n-1}(\nabla^{\alpha}R^{\mu\nu(l)})(\nabla^{\beta}R_{\mu\nu}^{(n-l-1)}),\quad \bar{\Omega}_{2}=\frac{1}{2M_{N}^{2}}\sum_{n=1}^{\infty}f_{2_{n}}\sum_{l=0}^{n-1}R^{\mu\nu(l)}R_{\mu\nu}^{(n-l)};\\
\nonumber\Delta^{\alpha\beta}_{2}&=&\frac{1}{4M_{N}^{2}}\sum_{n=1}^{\infty}f_{2_{n}}\sum_{l=0}^{n-1}\nabla^{\nu}\left(R_{\sigma\nu}^{(l)}\nabla^{(\alpha}R^{\beta)\sigma(n-l-1)}-(\nabla^{(\alpha}R_{\sigma\nu})R^{\beta)\sigma(n-l-1)}\right);\\
\nonumber \Omega^{\alpha\beta}_{3}&=&\frac{1}{2M_{N}^{2}}\sum_{n=1}^{\infty}f_{3_{n}}\sum_{l=0}^{n-1}(\nabla^{\alpha}C^{\mu(l)}_{\,\,\nu\rho\sigma})(\nabla^{\beta}C_{\mu}^{\,\,\nu\rho\sigma(n-l-1)}), \, \bar{\Omega}_{3}=\frac{1}{2M_{N}^{2}}\sum_{n=1}^{\infty}f_{3_{n}}\sum_{l=0}^{n-1}C^{\mu(l)}_{\,\nu\rho\sigma}C^{\,\,\nu\rho\sigma(n-l-1)}_{\mu};\\
\Delta^{\alpha\beta}_{3}&=&\frac{1}{4M_{N}^{2}}\sum_{n=1}^{\infty}f_{3_{n}}\sum_{l=0}^{n-1}\nabla^{\nu}\left(C_{\,\,\nu\sigma\mu}^{\rho(l)}\nabla^{(\alpha}C^{\,\,\beta)\sigma\mu(n-l-1)}_{\rho}-(\nabla^{(\alpha}C_{\,\,\nu\sigma\mu}^{|\rho(l)|})C^{\beta)\sigma\mu(n-l-1)}_{\rho}\right),
\end{eqnarray}
here we are using the same notation used in \cite{Biswas:2013cha}, i.e., $A^{(l)}\equiv \square^{l}A$. Of course the field equations are very cumbersome to solve since they are thoroughly nonlinear, then finding exact solutions does not seem to be a simple task. However, up to now exact solutions have been found for simplest models which, unlike the generic one (\ref{ac}), typically involved only one nonlocal term, namely the first one from (\ref{ac}) given by $R{\cal F}(\Box)R$. For example, it has also been shown in \cite{Modesto1}  that FRW and G\"{o}del metrics are exact solutions for the simplest model in which the form factors $\mathcal{F}_{2}$, $\mathcal{F}_{3}$ are taken {\it ad hoc} to be zero. In the next sections we shall examine the consistency of the class of ST-homogeneous G\"{o}del-type metrics as solutions within the non-local modified gravity theory.

\section{ST-homogeneous G\"{o}del-type metrics}  

This section is intended to give a brief review on the main features of a generalized class of metrics called ST-homogeneous G\"{o}del-type metrics. The ST-homogeneous G\"{o}del-type metrics are defined by the following line element in cylindrical coordinates \cite{Reb1}
\begin{equation}\label{type_godel}
    ds^2=[dt+H(r)d\theta]^2-D^{2}(r)d\theta^{2}-dr^{2}-dz^{2}, 
\end{equation}
where $H(r)$ and $D(r)$ are metric functions depending only on  radius coordinate $r$. Moreover, the homogeneity conditions in the space-time are attainable by the conditions
\begin{eqnarray}
& &\frac{H^{'}(r)}{D(r)}=2\omega,\\
& &\frac{D^{''}(r)}{D(r)}=m^{2},
\label{ST}
\end{eqnarray}
where the prime means derivative with respect to the radius coordinate. The constant metric parameters $(m^2, \omega)$ are restricted to take on values in the range: $-\infty \leq m^2 \leq \infty$ and $\omega\neq0$ (which is physically related to the vorticity of the space-time). As discussed in \cite{Reb1}, the ST-homogeneous G\"{o}del-type spaces can be separated into four different classes by depending on the $\omega$ value and the sign of $m^2$:
\begin{itemize}
\item \textit{hyperbolic class}: $m^2>0$, $\omega\neq 0$:
\begin{eqnarray}
&&H(r)=\frac{2\omega}{m^2}[\cosh(mr)-1],\\
&&D(r)=\frac{1}{m}\sinh(mr),
\end{eqnarray}
\item \textit{trigonometric class}: $-\mu^2=m^2<0$, $\omega\neq 0$:
\begin{eqnarray}
&&H(r)=\frac{2\omega}{\mu^2}[1-\cos(\mu r)],\\
&&D(r)=\frac{1}{\mu}\sin(\mu r),
\end{eqnarray}
\item \textit{linear class}: $m^2=0$, $\omega\neq 0$:
\begin{eqnarray}
&&H(r)=\omega r^2,\\
&&D(r)=r.
\label{linear}
\end{eqnarray}
\item \textit{degenerate class}: $m^2\neq 0$, $\omega=0$:
\begin{eqnarray}
H(r)=cte,
\label{degenerate}
\end{eqnarray}
\end{itemize}
in the degenerate case there is no rotation term in the metric since the function $H(r)$ can be conveniently chosen to vanish by means of a suitable coordinate transformation. The function $D(r)$ assumes the aforementioned forms by depending on the sign of $m^2$. 

 Note that the G\"{o}del metric itself \cite{Godel} is a particular example of the ST-homogeneous G\"{o}del-type spaces corresponding to  $m^2 =2\omega^2$, then it belongs to the hyperbolic class. In addition, another important feature of the ST-homogeneous G\"{o}del-type spaces concerns to the isometry group, for example, the class $m^2=4\omega^2$ admits the larger isometric group, $G_7$ \cite{Reb1}, whilst for $m^2<4\omega^2$ admits $G_5$ as the isometry group and the degenerate class present isometry group $G_6$.

The ST-homogeneous G\"{o}del-type spaces present Closed Time-like Curves (CTC's) which are circles $C=\lbrace(t,r,\theta,z); \, t, r, z= \mbox{const}, \theta \in [0, 2\pi]\rbrace$, defined in a region limited by the range ($r_1 <r<r_2$), where $G(r)=D^2 (r)-H^2 (r)$ becomes negative within this range. It is interesting to note that there is no CTC's for the hyperbolic class corresponding to $m^2\geq 4\omega^2$, otherwise, they can exist. Hence, for the range of parameters $0<m^2<4\omega^2$ there are possible CTC's inside the region corresponding to $r>r_{c}$, where $r_{c}$ is the critical radius (limiting radius separating the causal and non-causal regions) given by
\begin{equation}
\sinh^2\bigg(\frac{m r_{c}}{2}\bigg)=\bigg(\frac{4\omega^2}{m^2}-1 \bigg)^{-1}.
\label{rc}
\end{equation}  
Similarly, the linear and trigonometric classes also exhibit CTCs. Both cases display a non-causal region, namely: for the linear one, this region is hit for $r>r_{c}$ and the critical radius $r_{c}=\frac{1}{\omega}$. In the trigonometric case, the situation is more subtle since there exists an infinite set of alternating non-causal and causal regions (this is confirmed by the explicit form of $r_{c}$ in this case which is described by the equation similar to (\ref{rc}) but with the usual sine instead of the hyperbolic one). There is no CTCs in the degenerate class due to the fact that $G(r)$ is always positive for all values of $r$.  In the next section, we shall check the consistency of ST-homogeneous G\"{o}del-type metrics and also their causality properties inside the our non-local theory.

\section{ST-homogeneous G\"{o}del-type metrics within non-local gravity}

Our aim in this section is to check the consistency of G\"{o}del-type metrics in the non-local modified gravity (\ref{ac}). Before further proceeding, let us implement the matter sources. Here, we assume that the matter content is composed of the same components that in the GR case, it is a natural and reasonable choice to follow since we want to recover GR at some point. We will not explicit the matter sources because it will not be useful for us in what follows (see \cite{Reb,ourgodel} for the exact form of the energy-momentum tensor).   

 The next step is to implement the geometrical ingredients, i.e., the left-hand side of Eq. (\ref{ac1}). At this time, it is useful to define a set of complex null frames $\theta^{a}=e^{a}_{\,\,\,\mu}dx^{\mu}$. Here, we consider the NP formalism \cite{penrose1984spinors,chandrasekhar1998mathematical} where the tetrad is identified as $e^{a}_{\,\,\mu}\equiv\{l_{\mu},n_{\mu},m_{\mu},\bar{m}_{\mu}\}$ where  $l_{\mu}$ and $n_{\mu}$ are real null vectors and $(m_{\mu},\bar{m}_{\mu})$ is a pair of  mutually conjugated null vectors satisfying the following conditions 
\begin{equation} 
l_{\mu}n^{\mu}=1,\, m_{\mu}\bar{m}^{\mu}=-1,\, l_{\mu}l^{\mu}=n_{\mu}n^{\mu}=n_{\mu}m^{\mu}=n_{\mu}\bar{m}^{\mu}=l_{\mu}m^{\mu}=l_{\mu}\bar{m}^{\mu}=0,
\end{equation}
leading to metric decomposition
\begin{equation}
g_{\mu\nu}=2l_{(\mu}n_{\nu)}-2m_{(\mu}\bar{m}_{\nu)},
\end{equation}
here lower Latin letters label local tetrad indices. A good null tetrad basis choice for the metric (\ref{type_godel}) is:
  \begin{eqnarray}
  \label{nullframe}
	l_{\mu}&=&\frac{1}{\sqrt{2}}[1,0,H(r),1];\\
	n_{\nu}&=&\frac{1}{\sqrt{2}}[1,0,H(r),-1];\\
	m_{\mu}&=&\frac{1}{\sqrt{2}}[0,-i,D(r),0];\\
	\bar{m}_{\mu}&=&\frac{1}{\sqrt{2}}[0,i,D(r),0],
	\end{eqnarray}
yielding
\begin{eqnarray}
\theta^{(0)}&=&\frac{1}{\sqrt{2}}(dt+H(r)d\theta+dz);\\
\theta^{(1)}&=&\frac{1}{\sqrt{2}}(dt+H(r)d\theta-dz);\\
\theta^{(2)}&=&\frac{1}{\sqrt{2}}(D(r)d\theta-idr);\\
\theta^{(3)}&=&\frac{1}{\sqrt{2}}(D(r)d\theta-idr).
\label{null}
\end{eqnarray}

 In this frame, the field equations (\ref{ac}) read
\begin{eqnarray}
\nonumber E^{ab}&=&G^{ab}+\Lambda_{\mbox{cc}} \eta^{ab}+P_{1}^{ab}+P_{2}^{ab}+P_{3}^{ab}-2\Omega_{1}^{ab}+\eta^{ab}(\eta_{cd}\Omega^{cd}_{1}+\bar{\Omega}_{1})-2\Omega^{ab}_{2}+\\
\nonumber&+&\eta^{ab}(\eta_{cd}\Omega^{cd}_{2}+\bar{\Omega}_{2})-4\Delta^{ab}_{2}-2\Omega^{ab}_{3}+\eta^{ab}(\eta_{cd}\Omega^{cd}_{3}+\bar{\Omega}_{3})-8\Delta^{ab}_{3}=\\
&=&\kappa^{2}T^{ab},
\label{eq1}
\end{eqnarray} 
where the transformation rule to link tensorial representations in both frames is $A^{ab}=e^{a}_{\,\,\,\mu}e^{b}_{\,\,\,\nu}A^{\mu\nu}$. 

Before proceeding with the general model with all non-null form factors, let us deem the simplest reduced model: 
\begin{equation}
S=\frac{1}{2\kappa^2}\int d^{4}x\sqrt{-g}\, \bigg[R-2\Lambda_{\mbox{cc}}+\frac{1}{M^{2}_{N}}\bigg(R\mathcal{F}_{1}(\square)R\bigg)\bigg]+S_{m}[g_{\mu\nu},\Psi],
\label{ds}
\end{equation}
which is reached from the full theory just setting $\mathcal{F}_{2}=\mathcal{F}_{3}=0$. The field equations for this model is obtained by getting rid of all tensorial quantities with indices $(2)$ and $(3)$ in Eq.(\ref{ac}). More simplifications can be carried out by taking note of the fact that ST-homogeneous G\"{o}del-type metrics have constant Ricci scalar, more precisely, $R=2(m^2-\omega^2)$, as a result, the $n^{th}$ derivative acting on $R$ vanishes for all $n\geq 1$, i.e., $R\square^n R=0$, then the quantity $P_{1}^{\alpha\beta}$ dramatically simplifies while $\Omega_{1}^{\alpha\beta}$ and $\bar{\Omega}_{1}$ vanish throughout. To be more clear by using the form factor expansion (\ref{fr}), one can find      
\begin{eqnarray}
 P_{1}^{\alpha\beta}=\frac{f_{1_{0}}}{2M_{N}^{2}}\left[\left(4G^{\alpha\beta}R+g^{\alpha\beta}R^2\right)\right],
\end{eqnarray}
where $f_{1_{0}}$ is the zeroth-order coefficient in the power series of the form factor $\mathcal{F}_{1}$. Then, there exist two possibilities: $1)$ $f_{1_{0}}=0$, it implies $P_{1}^{\alpha\beta}=0$, thereby the field equations recover GR ones. Therefore, in this particular case, ST-homogeneous G\"{o}del-type metrics are solutions both in GR and in the non-local model (\ref{ds}). In particular, the solution correspondent to G\"{o}del metric ($m^{2}=2\omega^2$) has been found previously in \cite{Modesto1}. $2)$ $f_{1_{0}}\neq 0$, it implies $P_{1}^{\alpha\beta}\neq 0$ for generic values of the metric parameters. This case turns out to reproduce the same conclusions to the $f(R)=R+\alpha R^2$ gravity since the derivatives of Ricci scalar play no relevant role in the field equations. In particular, when $m^2=\omega^2$, the metric is Ricci-flat so that $P_{1}^{\alpha\beta}=0$ regardless of particular choices of the form factor, thereby the field equations recover the GR ones again. Thenceforth, we will throw away the zeroth-order coefficient in the power series expansion of the form factors because of the fact that it only plays an important role in the UV regime, and since we are interested in the classical field equations (IR regime) it can be safely set to be zero as pointed out in \cite{Modesto1}.

\subsection{ST-homogeneous G\"{o}del-type metrics as a space with all constant scalar invariants (CSI)}

The CSI spaces are those whose all scalar curvature invariants ($\mathcal{I}$) are constants \cite{Coley:2005sq,Coley:2008th}. We mean by scalar invariants the set of all invariants made up from the curvature tensors and their derivatives, i.e.,
\begin{equation}
\mathcal{I}=\{R, R_{\mu\nu}R^{\mu\nu},C_{\mu\nu\alpha}C^{\mu\nu\alpha}, \nabla_{\alpha}R_{\mu\nu\beta\sigma}\nabla^{\alpha}R^{\mu\nu\beta\sigma},...\}.
\end{equation}    

Many theorems with important properties of CSI spaces has been proven in various works \cite{Coley:2008th,Coley:2009eb}. One of them will be relevant for us, namely:

\begin{itemize}
	\item \textbf{Theorem}: A 4-dimensional CSI spacetime is either:
	
	1) locally homogeneous; or
	
	2) a subclass of Kundt spacetimes.
\end{itemize}

From the above theorem, ST-homogeneous G\"{o}del-type metrics belong to the class of CSI spaces.

 Furthermore, according to the weight decomposition the algebraic classification of curvature (see Appendix \ref{appendix2}) of such metrics are Petrov (Weyl) type D or simpler which then implies that the Weyl scalars (in NP formalism) take the form:
\begin{equation}
\Psi_2=\frac{1}{6}(4\omega^2 - m^2),\quad \Psi_0=\Psi_1=\Psi_3=\Psi_4=0,
\end{equation} 
with $\Psi_2$ being constant as a consequence of the local homogeneity of the space. Note that the special class $m^2=4\omega^2$, in particular, is Petrov type 0, as a result it is conformally flat since all Weyl scalars vanish.
 
 For Petrov type D metrics, the only quadratic Weyl invariant $I=\frac{1}{2}\mathcal{C}_{\mu\nu\alpha\beta}\mathcal{C}^{\mu\nu\alpha\beta}$ constructed from the Riemann curvature tensor \cite{alcubierre2008introduction} is 
\begin{equation}
I=3\Psi^{2}_{2}=\frac{1}{12}(4\omega^2 -m^2)^2,
\label{Ij}
\end{equation}	
where $\mathcal{C}_{\mu\nu\alpha\beta}=\frac{1}{4}\left(C_{\mu\nu\alpha\beta}-iC^{\star}_{\mu\nu\alpha\beta}\right)$ with  $C^{\star}_{\mu\nu\alpha\beta}$ being the dual Weyl tensor. Due to the symmetries of the ST-homogeneous G\"{o}del-type metrics the total contraction between the Weyl tensor and its dual one is zero, i.e., $C^{\mu\nu\alpha\beta}C^{\star}_{\mu\nu\alpha\beta}=0$, thereby from Eq. (\ref{Ij}) we have
\begin{equation}
C_{\mu\nu\alpha\beta}C^{\mu\nu\alpha\beta}=48\Psi^{2}_{2}=\frac{4}{3}(4\omega^2 -m^2)^2.
\end{equation}

In respect the Ricci tensor in the complex null frame (\ref{null}), the non-vanishing components are
\begin{eqnarray}
R_{(0)(0)}&=&2\Phi_{00}=\omega^2;\\
R_{(1)(1)}&=&2\Phi_{22}=\omega^2;\\
R_{(0)(1)}&=&2\left(\Phi_{11}-3\Lambda\right)=\omega^2;\\
R_{(2)(3)}&=&2\left(\Phi_{11}+3\Lambda\right)=2\omega^2-m^2,
\end{eqnarray}
where the Ricci scalars $\Phi_{AB}$ and $\Lambda$ are defined in \cite{alcubierre2008introduction}.

\subsection{Completely causal solutions in non-local gravity}

We now turn our attention back to the complete theory (\ref{ac}). As it was said before, ST-homogeneous G\"{o}del-type metrics are locally homogeneous, with all corresponding scalar curvature invariants are constant. Such properties mean that it is always possible to pick a local frame where the components of the curvature (or Weyl) tensor and their covariant derivatives take constant values. Keeping this in mind together with the previous information on the geometrical quantities, we are able to evaluate the higher order tensors. Therefore, let us start by evaluating the first-order derivative of the Weyl tensor which we have found that the non-vanishing components in the complex null basis take the following canonical form:
\begin{equation}
\nabla_{a}C_{bcde}\propto \omega(4\omega^2 -m^2).
\label{na}
\end{equation}
The non-vanishing components of the second- and third-order derivative of the Weyl tensor also can be rewritten in terms of the former canonical form as shown below
\begin{eqnarray}
\nabla_{a}\nabla_{b}C_{cdef}&\propto & \omega^{2}(4\omega^2 -m^2);\\
\nabla_{a}\nabla_{b}\nabla_{c}C_{defg} &\propto & \omega^{3}(4\omega^2 -m^2),
\label{nb}
\end{eqnarray}
we can keep doing this up to $n^{th}$-order and the canonical form will hold. 
This is not a surprise since it is just a consequence of the Cartan-Karlhede algorithm \cite{Karlhede:1979ri}. In fact, any other higher-order derivatives will yield the same canonical form given by Eqs. (\ref{nb}).

Through similar arguments, we find
\begin{eqnarray}
\nonumber\nabla_{a}R_{bc}&\propto& \omega(4\omega^2 -m^2);\\
\nabla_{a}\nabla_{b}R_{cd}&\propto& \omega^{2}(4\omega^2-m^2);\\
\nonumber&.&\\
\nonumber&.&\\
\nonumber&.&
\end{eqnarray}
With this at hands, one can construct the $n^{th}$-order invariants we are concerned of,
\begin{eqnarray}
C^{\mu\nu\alpha\beta}\square C_{\mu\nu\alpha\beta}&=&-8\omega^2 (4\omega^2 -m^2);\\
C^{\mu\nu\alpha\beta}\square^{2} C_{\mu\nu\alpha\beta}&=&48\omega^4 (4\omega^2 -m^2);\\
C^{\mu\nu\alpha\beta}\square^{3} C_{\mu\nu\alpha\beta}&=&-288\omega^6 (4\omega^2 -m^2);\\
&&.\nonumber\\
&&.\nonumber\\
&&.\nonumber\\
C^{\mu\nu\alpha\beta}\square^{n} C_{\mu\nu\alpha\beta}&=&(-1)^{n}6^{n-1}8\omega^{2n} (4\omega^2 - m^2),\,\,\mbox{for}\quad n\geq 1.
\end{eqnarray}  
Analogously,
\begin{eqnarray}
R^{\mu\nu}\square R_{\mu\nu}&=&-4\omega^{2}(4\omega^2 -m^2);\\
R^{\mu\nu}\square^{2}R_{\mu\nu}&=&24\omega^{4}(4\omega^2 -m^2);\\
R^{\mu\nu}\square^{3}R_{\mu\nu}&=&-144\omega^{6}(4\omega^2 -m^2);\\
&&.\nonumber\\
&&.\nonumber\\
&&.\nonumber\\
R^{\mu\nu}\square^{n}R_{\mu\nu}&=&(-1)^{n}6^{n-1}4\omega^{2n}(4\omega^2 -m^2),\,\,\mbox{for}\quad n\geq 1.
\end{eqnarray}
We recall that $R\square^{n}R=0$ for $n\geq 1$ as we already concluded.

 As far as the field equations (\ref{ac1}) are concerned, it turns out that the above argumentation can be applied for the other high-order derivative tensorial quantities of the field equations in the complex null frame (\ref{nullframe}), that is, $P_{i}^{ab}, \Omega_{i}^{ab}, \bar{\Omega}_{i}^{ab}$ and $\Delta_{i}^{ab}$ with $i=2,3$, are constants and they vanish for $\omega=0$ (degenerate class) and for $m^2=4\omega^2$. Otherwise, the field equations reduce to a set of algebraic equations with infinite degree in $\omega$ to be solved. Obviously, it establishes an overcomplete system of algebraic equations, in this case it  becomes undoubtedly impracticable to find solutions. Returning to the former case, i.e., $\omega=0$ and $m^2=4\omega^2$ are exact solutions of the non-local theory (\ref{actmod}) and the field equations reduce to the GR ones. Therefore, the only ST-homogeneous G\"{o}del-type solutions that hold in our non-local gravity theory are those classes characterized by $\omega=0$ with arbitrary $m\neq 0$ and  $m^2=4\omega^2$. It is noteworthy that both solutions are completely causal providing that the full non-local ghost-free (\ref{ac}) only support solutions avoiding CTC's unlike the model (\ref{ds}).

\section{Summary and conclusions}
\label{secIII}

We have investigated ST-homogeneous G\"{o}del-type metrics within the non-local gravity theory (\ref{ac}). Despite the highly non-linear form of field equations we have succeeded in engendering exact solutions for this model. It was first examined the particular model corresponding to the Eq. (\ref{ds})  wher the form factors $\mathcal{F}_{2}$ and $\mathcal{F}_{3}$ are put to zero. In this case all ST-homogeneous G\"{o}del-type metric classes are consistent within this particular model when, first, the first-order expansion coefficient of the form factor vanishes, then the field equations recover GR, and obviously their solutions are the same of GR, second, the first-order expansion coefficient of the form factor does not vanish, in this situation, the solutions are the same of a particular model $f(R)=R+\alpha R^2$. 

Regarding the whole theory (\ref{ac}) we have shown that the exact solutions found -- they correspond to $\omega=0$ and $m^2=4\omega^2$ -- display the remarkable property of being completely causal. Physically speaking, the unitary theory only ``selects" those ST-homogeneous G\"{o}del-type metrics circumventing the presence of CTCs by some mechanism coming from the incorporation of the infinite higher-derivative terms in the gravitational action. It suggests a strong link between ghost-free theories like Eq. (\ref{ac}) and the Chronological Protection principle. Of course, a more careful assessment regarding this point should be made. Such investigations are currently in progress.


\appendix

\section{Boost-weight decomposition}
\label{appendix2}

In this appendix we briefly discuss on boost-weight decomposition. First let us consider a complex null frame $\{l_{\mu},n_{\mu},m_{\mu},\bar{m}_{\mu}\}$ that satisfies the following conditions:
\begin{equation}
l_{\mu}n^{\mu}=1,\, m_{\mu}\bar{m}^{\mu}=-1,\, l_{\mu}l^{\mu}=n_{\mu}n^{\mu}=n_{\mu}m^{\mu}=n_{\mu}\bar{m}^{\mu}=l_{\mu}m^{\mu}=l_{\mu}\bar{m}^{\mu}=0.
\end{equation}
 We now apply a boost in the $(l_{\mu}-n_{\mu})$ plane, i.e.,
\begin{equation}
\left\{\tilde{l}_{\mu},\tilde{n}_{\mu},\tilde{m}_{\mu},\tilde{\bar{m}}_{\mu}\right\}=\left\{e^{\lambda}l_{\mu},e^{-\lambda}n_{\mu},m_{\mu},\bar{m}_{\mu}\right\}.
\label{kl}
\end{equation}
A generic tensor $T$ can be decomposed in terms of the boost weight with respect to the boost transformation (\ref{kl}) as follows:
\begin{equation}
T=\sum_{b} \left(T\right)_b,
\end{equation}
where $\left(T\right)_b$ stands for the projection of the tensor $T$ onto the vector space of boost-weight $b$. In respect the components of the tensor $T$ under the boost-weight transformation, they will transform as:
\begin{equation}
\left(T\right)_{b\,\,\mu\nu...}=e^{-b\lambda}\left(T\right)_{b\,\,\tilde{\mu}\tilde{\nu}...},
\end{equation}  
where indices with tildes label boosted tensorial components.

The boost-weight decomposition of the metric is simply: $g=(g)_{0}$, i.e., the metric has only boost weight $0$ component. Similarly, it easy to see  that any scalar invariant of a tensor, say $T$, should be boost-weight invariant since all scalar terms are invariant under the action of the $SO(1,3)$ group, thus:
\begin{equation}
\mbox{Contr}\left[T\right]=\mbox{Contr}\left[\left(T\right)_{0}\right],
\end{equation}
where ``Contr'' stands for full contraction.

Spacetimes can be algebraically classified based on boost-weight decomposition of Weyl tensor--the well-known Petrov classification in 4D, namely:
\begin{itemize}
	\item Type I: $C=\left(C\right)_{-2}+\left(C\right)_{-1}+\left(C\right)_{1}+\left(C\right)_{0}$;
	\item Type II: $C=\left(C\right)_{-2}+\left(C\right)_{-1}+\left(C\right)_{0}$;
	\item Type D: $C=\left(C\right)_{0}$;
	\item Type III: $C=\left(C\right)_{-2}+\left(C\right)_{-1}$;
	\item Type N: $C=\left(C\right)_{-2}$;
	\item Type O: $C=0$.
\end{itemize}
  The classification of the Weyl tensor in higher dimensions has been developed in \cite{Coley:2004jv}.

\acknowledgments
P. J. Porf\'irio would like to thank the Brazilian agency CAPES financial support (PDE/CAPES grant, process 88881.171759/2018-01). The work by A. Yu. P. has been partially supported by CNPq, project 301562/2019-9.


\end{document}